\documentclass[a4paper]{jpconf}
\usepackage{graphicx}
\usepackage[square,sort&compress,numbers]{natbib}
\usepackage{subfigure}
\begin{document}
\title{GPU Enhancement of the Trigger to Extend Physics Reach at the LHC}

\author{P Lujan, V Halyo, A Hunt, P Jindal, P LeGresley}

\address{Department of Physics, Princeton University, Princeton, NJ 08544, USA}

\ead{plujan@princeton.edu}

\begin{abstract}
At the Large Hadron Collider (LHC), the trigger systems for the detectors must be able to process a very large
amount of data in a very limited amount of time, so that the nominal collision rate of 40 MHz can be reduced
to a data rate that can be stored and processed in a reasonable amount of time. This need for high performance
places very stringent requirements on the complexity of the algorithms that can be used for identifying events
of interest in the trigger system, which potentially limits the ability to trigger on signatures of various
new physics models. In this paper, we present an alternative tracking algorithm, based on the Hough transform,
which avoids many of the problems associated with the standard combinatorial track finding currently used. The
Hough transform is also well-adapted for Graphics Processing Unit (GPU)-based computing, and such GPU-based
systems could be easily integrated into the existing High-Level Trigger (HLT). This algorithm offers the
ability to trigger on topological signatures of new physics currently not practical to reconstruct, such as
events with jets or black holes significantly displaced from the primary vertex. This paper presents, for the
first time, an implementation and preliminary performance results using NVIDIA Tesla C2075 and K20c GPUs.
\end{abstract}

\section{Introduction}
In a modern high-energy physics experiment such as the CMS or ATLAS detectors at the Large Hadron Collider
(LHC), the trigger system is one of the critical components, as it must process the raw data coming from the
detector at the nominal LHC collision rate of 40 MHz and identify events of interest in real time, in order to
reduce the data rate to one within the capabilities of the offline storage and event reconstruction
systems. This is achieved in CMS and ATLAS by means of a two-tiered trigger system, with a Level 1 (L1)
trigger implemented in hardware and firmware which uses fast and relatively simple algorithms to quickly
identify events for further processing, and then a high-level trigger (HLT) which uses a farm of commercial
CPU processors to reconstruct the event in further detail, using algorithms similar to those used in the
offline reconstruction to select events containing physical processes of interest. The following discussion is
based on experience with the CMS HLT~\cite{Sakulin:2007rj}, but is applicable to any similar experiment.

The design of HLT algorithms is thus defined by the balance of two competing needs: the goal of achieving the
best possible selection of interesting physics events, and the need to execute that selection in a very
limited amount of processing time. This means that algorithms which are relatively computationally expensive
are not practical at the HLT. An example of how this limits the physics abilities can be seen in the track
reconstruction performed at the HLT. In CMS, this proceeds using the standard Combinatorial Track Finder (CTF)
algorithm, a simple and well-understood algorithm. The combinatorial nature of this algorithm, however, means
that as the luminosity and hence the number of hits in the detector increases, the number of possible
combinations and so the running time of the algorithm increases considerably. In order to limit the CPU
usage at the HLT, tracks which are significantly displaced from the primary interaction point are excluded
from consideration. This saves processing time, but at the cost of potentially missing new physics events with
this signature.

Improving the performance of event processing at the HLT thus opens up the potential for extending the physics
reach of the LHC by making it possible to trigger on signatures not previously practical. One promising avenue
for realizing this improvement is by integrating Graphics Processing Unit (GPU)-based computing, using
technologies such as the NVIDIA Tesla line, designed to be programmable with general-purpose programming
techniques for high-performance computing. The flexible nature of the HLT processor farm means that such new
technologies could be easily integrated. Such additions would also make possible the introduction of entirely
new algorithms which could reconstruct the event more efficiently than currently possible. In this paper, we
present an alternative tracking algorithm based on the Hough transform, and show an implementation of this
algorithm tested on two different NVIDIA Tesla cards. This algorithm offers the possibility of triggering on a
wide variety of new physics models which are not identified by the current CMS trigger.

\section{Physics motivation}

The topological signature of tracks (either from jets or leptons) originating from a secondary vertex at a
significant displacement from the primary interaction point would represent a clear and distinct signal of new
physics beyond the Standard Model (SM). A wide variety of models predicts such a
signature~\cite{Halyo:2013yfa}; examples include hidden valley models with long-lived neutral particles,
R-parity-violating supersymmetric models with long-lived neutralinos, displaced black holes, boosted jets, or
Z$^{\prime}$ models with long-lived neutrinos.

While searches have been performed at ATLAS and CMS for some of these signatures, they are limited by the
constraints of the trigger system, preventing us from searching in many regions of interest. For example, CMS
has performed a search for a (non-Standard Model) Higgs boson producing long-lived neutral particles which
decay into displaced leptons~\cite{Chatrchyan:2012jna}, but for a mass close to the SM Higgs mass of $125$
GeV$/c^2$, the trigger efficiency is extremely low due to the lepton momentum requirements in the trigger,
resulting in relatively poor mass limits. A trigger with the ability to identify tracks originating from a
significantly displaced vertex would thus open the possibility to search in a previously inaccessible region
of parameter space for this model.

\section{Tracking algorithms}

Track finding in the CMS HLT (as in the offline reconstruction) proceeds using the Combinatorial Track Finder
(CTF) algorithm. This algorithm is an iterative process in which earlier iterations search for tracks which
can be relatively easily reconstructed (e.g., higher-momentum tracks); the hits in these tracks are then
removed, allowing later iterations to search for more difficult-to-reconstruct tracks without the number of
combinatorial possibilities becoming too large. In the HLT, the CTF is run only regionally; that is, rather
than reconstructing tracks in the whole detector, only regions of interest around previously-identified
objects (such as energy deposits in the calorimeters or tracks in the muon chambers) are reconstructed.

In each iteration, initial trajectories are constructed from seeds consisting of either a triplet of hits or a
pair of hits and a beamspot constraint. The trajectories are then propagated through the detector, and at each
layer of the detector, hits compatible with the trajectory are searched for and, if found, attached and fit to
the trajectory. Because of uncertainties in the track parameters and the hit positions, a trajectory may be
compatible with more than one hit in a layer; in this case, new candidates are constructed for each
possibility and propagated through the rest of the detector separately. The track candidates are then cleaned
to select only the best candidate from these possibilities and eliminate duplicate tracks. After this step, a
full Kalman fit is performed to obtain the best estimate of the track parameters at all points on the
trajectory. This fit uses a Runge-Kutta propagator to take account of material effects and inhomogeneities in
the magnetic field. Finally, selection requirements are applied to obtain a sample of high-quality tracks.

In the CMS offline reconstruction, the last iteration is designed to reconstruct highly displaced tracks, and
uses seeds from the outer layers of the tracker with relatively large values of the track impact parameter
allowed. However, because this step is computationally expensive, no equivalent step is run at the HLT level;
as a result, no tracks with a transverse impact parameter (the transverse distance to the primary interaction
at the point of closest approach) greater than 1.0 cm are reconstructed in the CMS HLT.

The Hough transform, in contrast, represents an entirely different approach to the problem of track
reconstruction. Instead of reconstructing tracks individually, it can reconstruct all tracks in the event in a
single pass. The Hough transform is an image processing algorithm for feature detection that considers all
possible instances of a parameterized feature, such as a line or circle.  Each possible instance of a feature
starts with zero votes in the parameter space, and then, for each piece of input data, votes are added to the
feature instances that would include that input data.  After all input data has been processed, the votes in
the parameter space are examined.  Locations in the parameter space with more votes are likely to be actual
features in the input data, so this step amounts to looking for local maxima in the parameter space. After the
initial identification, further computations (e.g., a full Kalman fit) can be performed to confirm the
existence of the track and obtain the best measurement of the track parameters.

\begin{figure}[!Hhtb]
\begin{center}
	\subfigure[]{\includegraphics[width=0.32\textwidth]{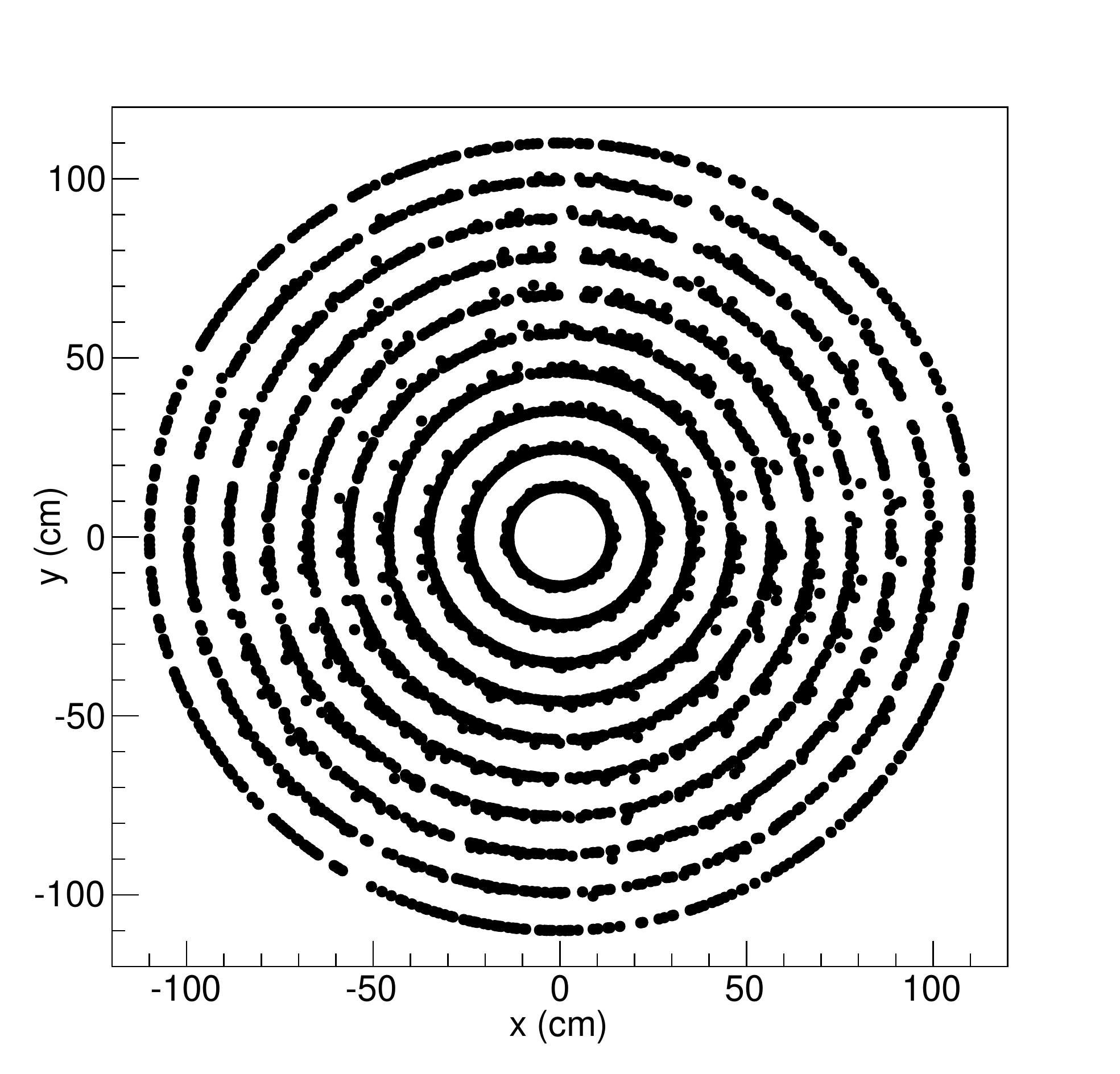} \label{fig:hits}}
	\subfigure[]{\includegraphics[width=0.30\textwidth]{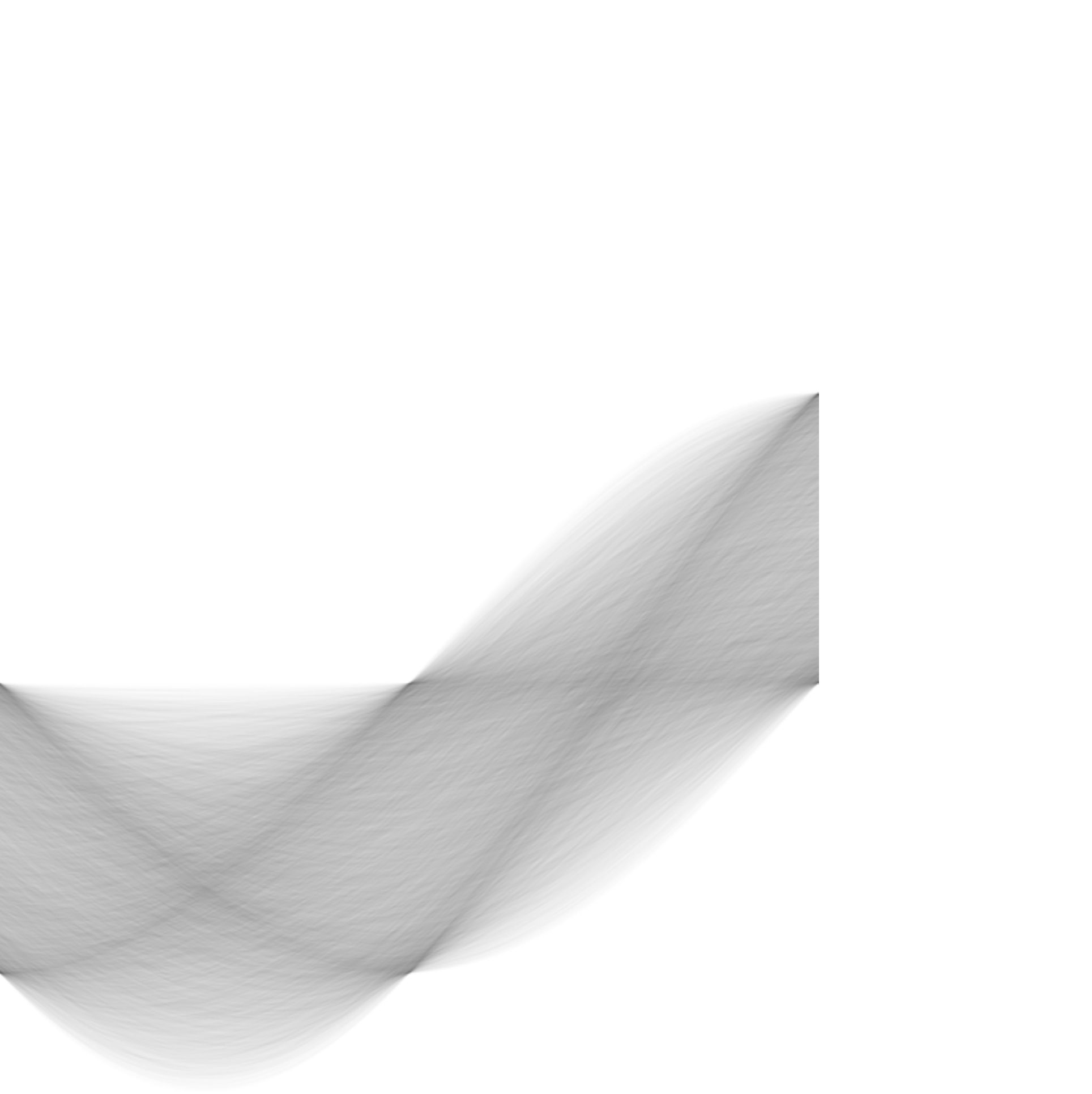} \label{fig:accumulator}}
	\subfigure[]{\includegraphics[width=0.32\textwidth]{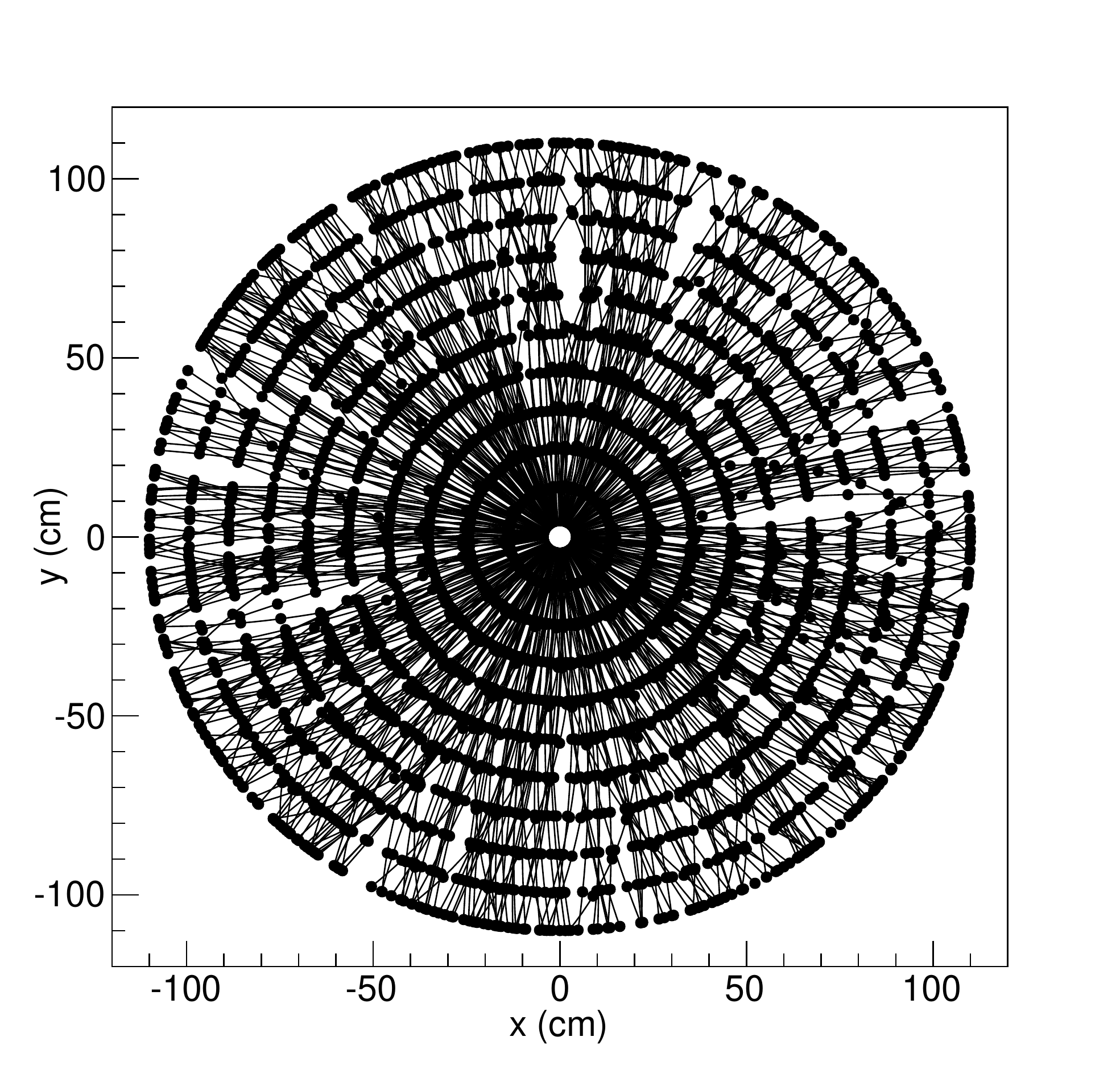} \label{fig:tracks}}
	\caption{The Hough transform algorithm applied to a case with curved tracks. Left: Hits from Monte Carlo
	simulation for curved tracks originating from the interaction point. Center: The parameter space
	obtained after applying the Hough transform. Right: Final reconstructed tracks. The resulting
	track-finding efficiency for this example is approximately 86\%.\label{fig:hough_curved}}
\end{center}
\end{figure}

Figure~\ref{fig:hough_curved} shows an example of the algorithm in operation. In this simulation, a simple
detector model considering only the transverse plane is used. The simulated detector has a beampipe with
radius of 3.0~cm, surrounded by ten concentric, evenly-spaced tracking layers with an outer radius of
110.0~cm. The resolution for a single hit is taken to be 0.4~mm in each direction. In Figure~\ref{fig:hits},
we see the simulated hits for 500 simulated curved tracks. The Hough transform is then applied, resulting in
the parameter space shown in Figure~\ref{fig:accumulator}. The maxima in the parameter space are then
identified, corresponding to the reconstructed tracks illustrated in Figure~\ref{fig:tracks}.

The Hough transform offers considerable advantages over the traditional CTF algorithm as the number of hits in
the detector increases, as the Hough transform running time is dependent on the number of hits, while the
number of combinations that need to be considered in the CTF increases at a faster rate. Thus, the Hough
transform algorithm is well-positioned to deal with the high-pileup environment expected in further LHC
operations. It should be noted, however, that the performance of the Hough transform is highly dependent on
the dimensionality of the parameter space. The two-parameter space presented here can identify curved tracks
passing through the origin, or straight displaced tracks, but curved, displaced tracks require three
parameters to describe. This has not yet been implemented and optimized.

\section{GPU implementation}

With recent interest in GPUs as an additional source of highly parallel computing power, technologies have
been introduced, such as NVIDIA's CUDA, to enable usage of GPUs for general-purpose computing using nearly
standard C/C++ programming, and the NVIDIA Tesla line is a line of GPUs designed specifically for this
role. In this paper we consider two Tesla models: the C2075, which features 448 cores at a speed of 1.15 GHz
and 6 GB of onboard RAM, and the K20c, which has 2496 cores at 706 MHz and 5 GB of RAM.

\begin{figure}[!Hhtb]
\begin{center}
\includegraphics[width=0.5\textwidth]{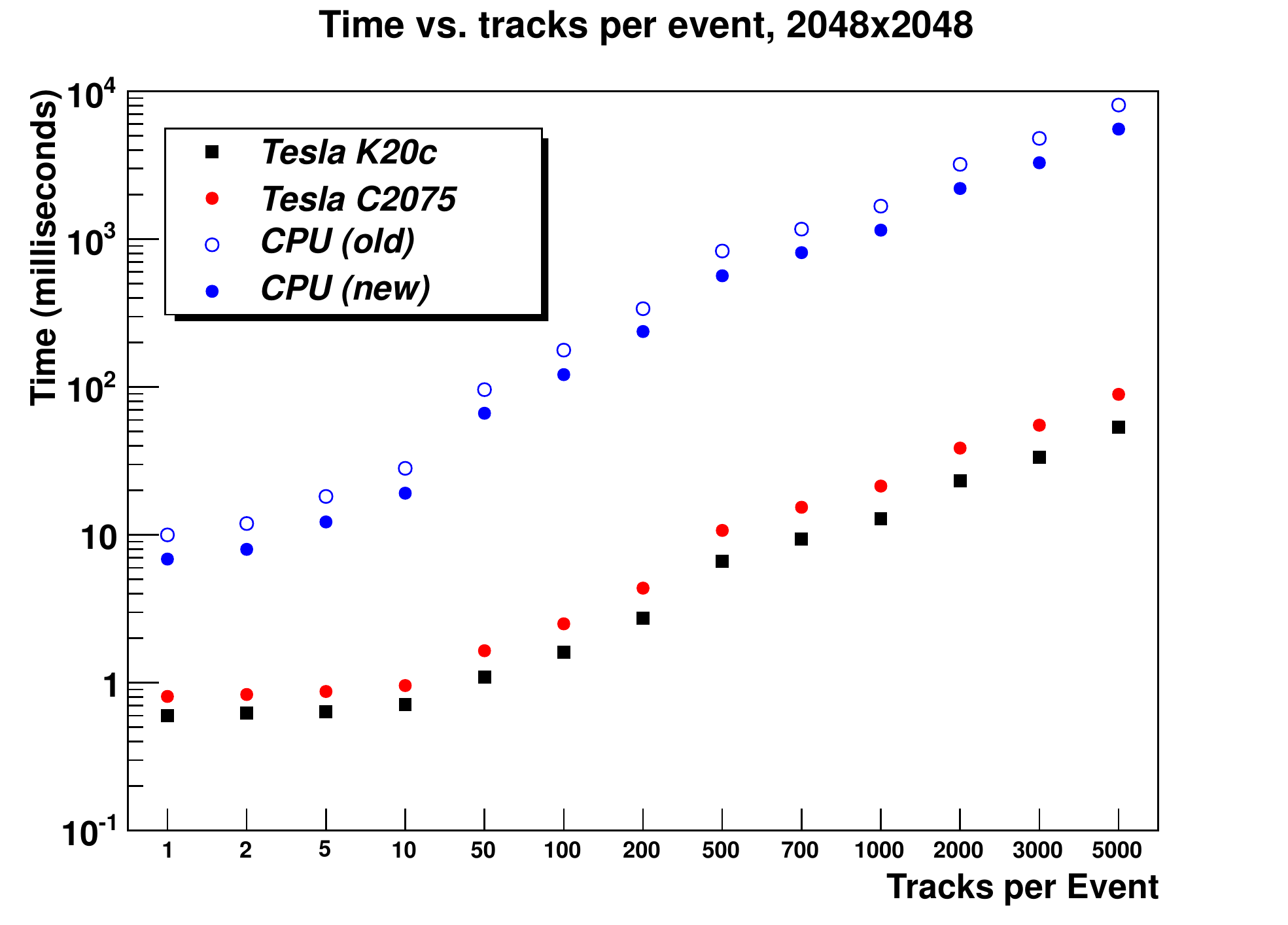} 
\caption{Time performance, as a function of number of tracks in the event, of our CPU implementation of the
 Hough transform before (old) and after (new) optimization, running on an Intel Core i7-3770 CPU, compared to
 the GPU implementation running on an NVIDIA Tesla C2075 and K20c.
\label{fig:TimePerformance}}
\end{center}
\end{figure}

The Hough transform is naturally amenable to a high degree of parallelization, as the parameter space
calculation for each hit is independent of all other hits in the event. Consequently, it is a natural
candidate for implementation on a GPU. In order to provide a baseline for comparison, we also built a
multithreaded CPU-based implementation of the Hough transform, which was run on an Intel Core i7-3770 (Ivy
Bridge) CPU. After the initial CPU implementation was tested, additional optimizations were performed to
improve the multithreading performance. Figure~\ref{fig:TimePerformance} shows the performance of the
resulting implementations~\cite{Halyo:2013iba,Halyo:2013cza}. The GPU implementation is seen to perform
significantly better even after the CPU implementation has been optimized.

\section{Displaced vertex identification}

\begin{figure}[!Hhtb]
\begin{center}
  \subfigure[]{\includegraphics[width=0.35\textwidth]{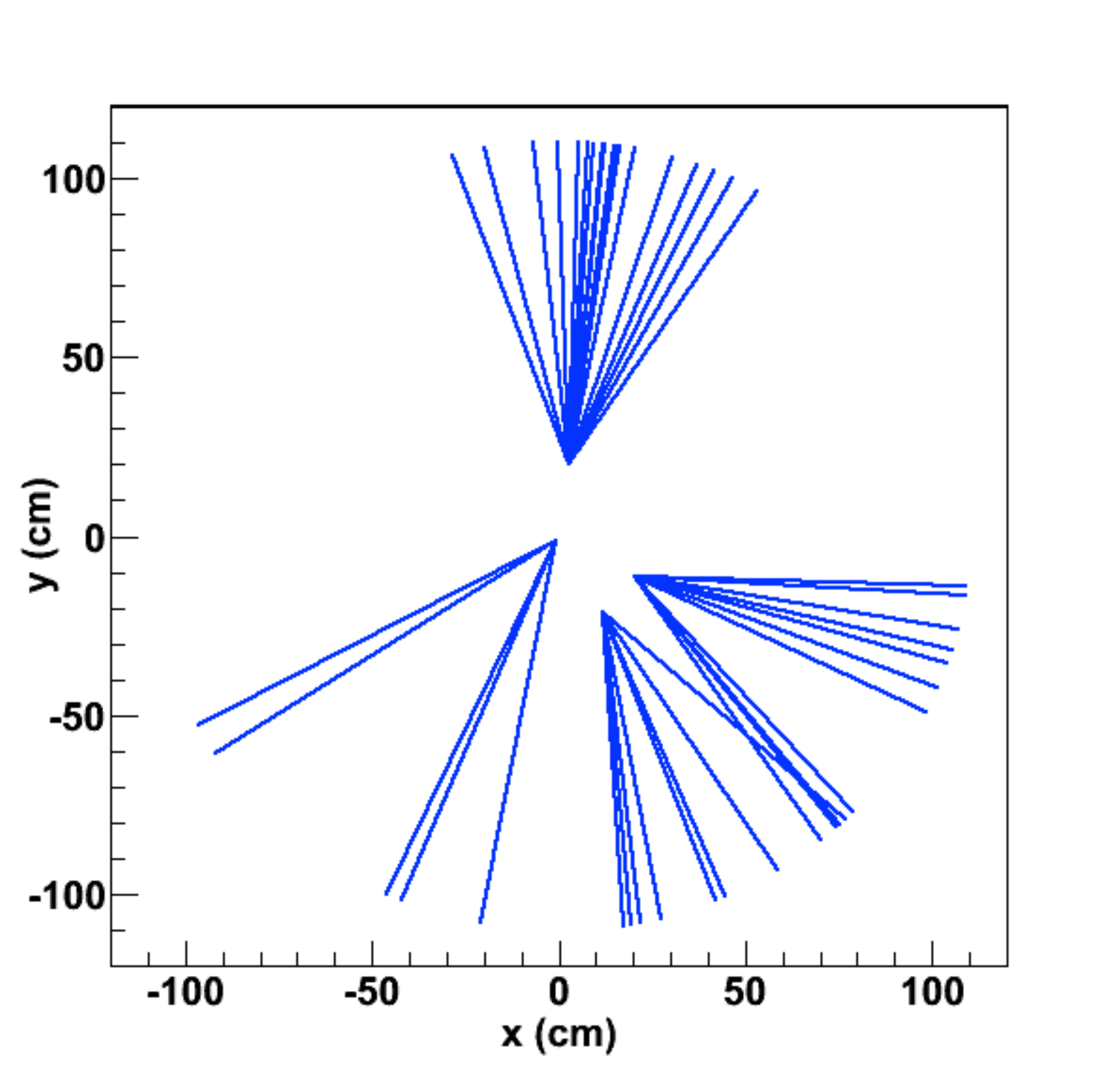} \label{fig:DisplacedJetsTracks}}
  \subfigure[]{\includegraphics[width=0.30\textwidth]{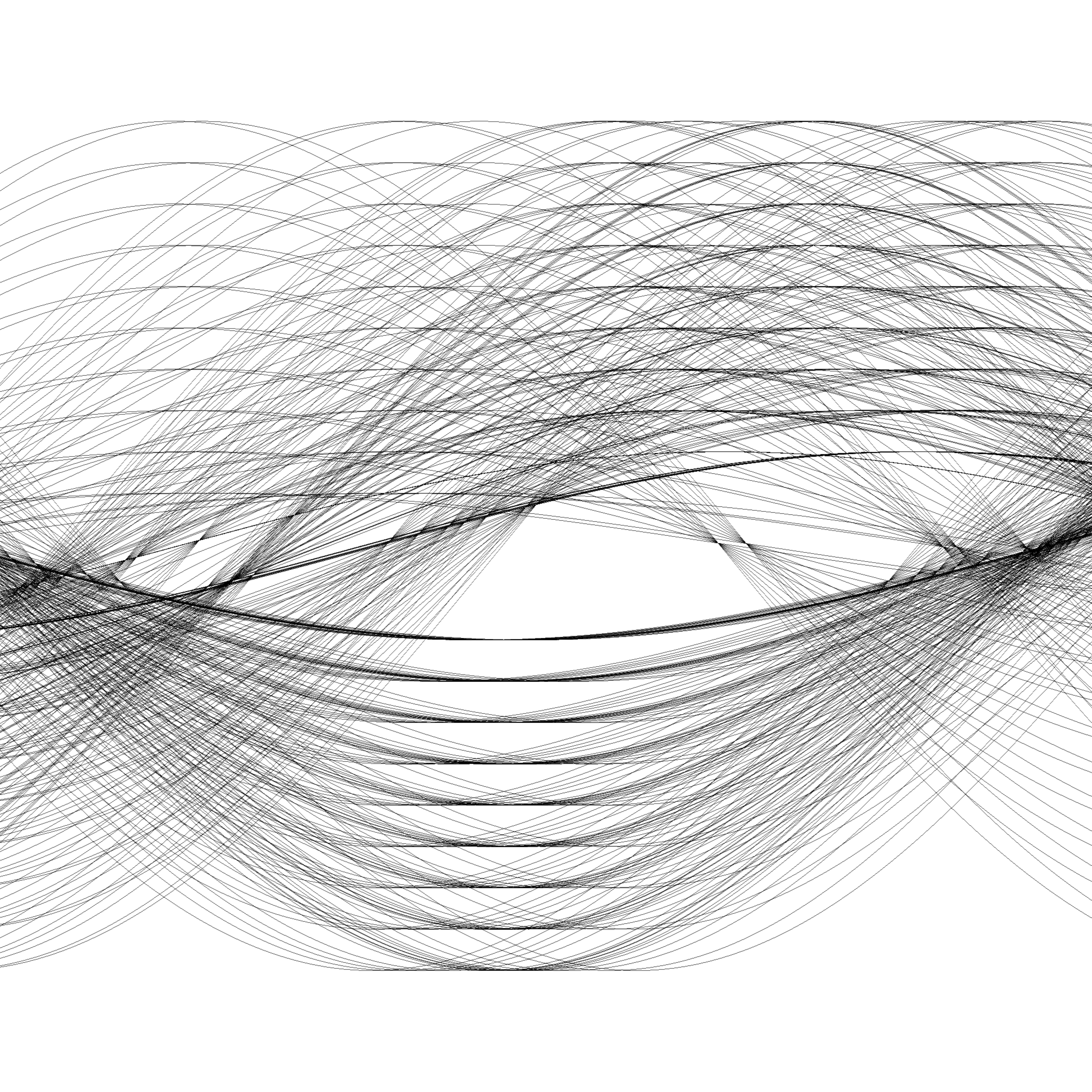} \label{fig:DisplacedJetsAccumulator}}
  \subfigure[]{\includegraphics[width=0.30\textwidth]{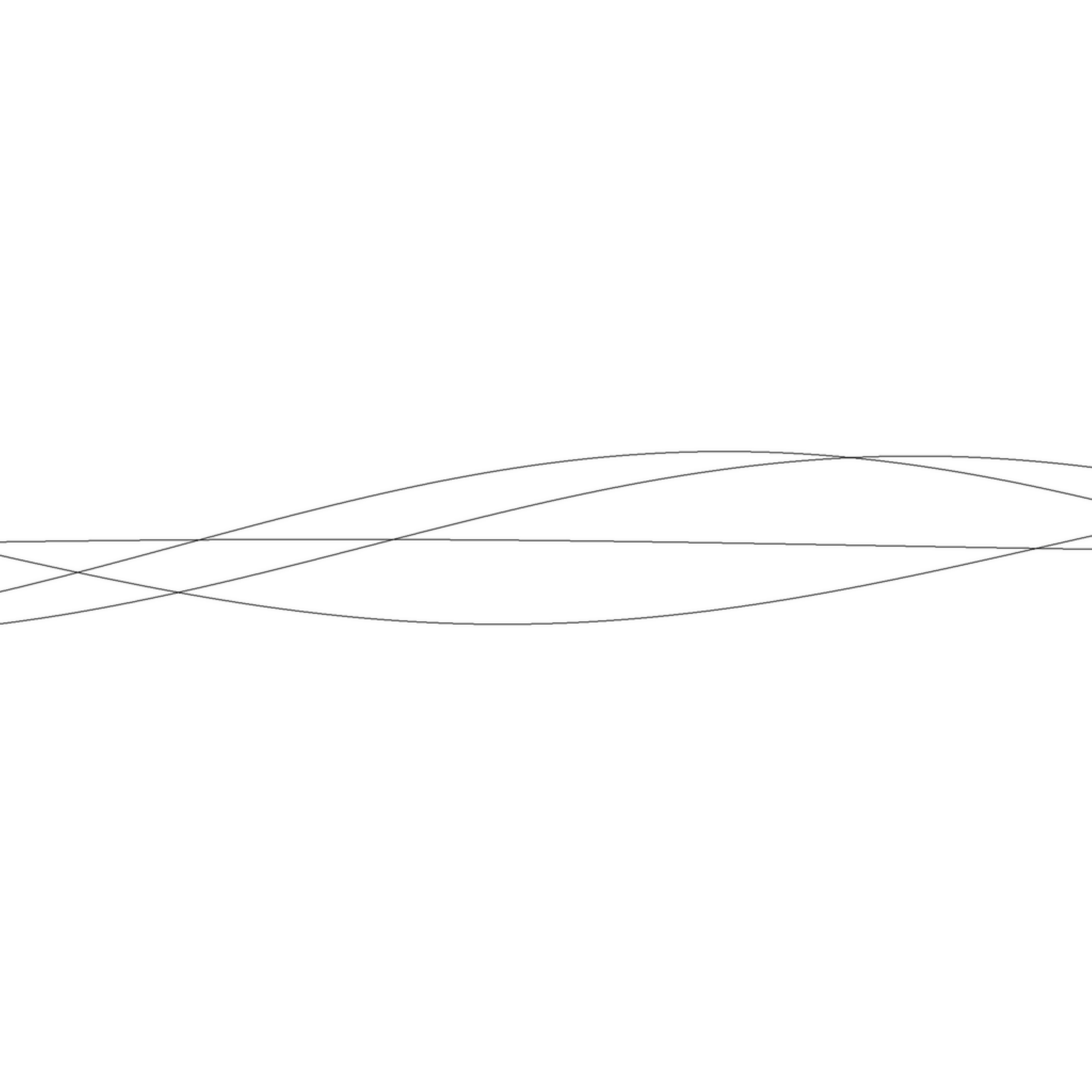} \label{fig:DisplacedJetsVertex}}
  \caption{Hough transform algorithm applied to an event with multiple displaced jets. Left: Simulated tracks
    in the event. Center: The first Hough transform identifies the tracks present in the event. Right: The
    second Hough transform identifies the sinusoids corresponding to the jet vertices.
    \label{fig:DisplacedJets}}
\end{center}
\end{figure}

In addition to displaced tracks, the ability to identify displaced vertices would be particularly useful for
models with long-lived neutral particles, as this would be an especially clear signal of new
physics. Figure~\ref{fig:DisplacedJets} shows how this vertex identification can be simply achieved using the
Hough transform. In Figure~\ref{fig:DisplacedJetsTracks}, we see simulated tracks for an event with four jets,
each of which has a vertex significantly displaced from the origin. Figure~\ref{fig:DisplacedJetsAccumulator}
shows the results of the Hough transform applied to the resulting hits, where the tracks are the maxima points
in the parameter space. The vertices can be now be found by applying a second Hough transform to the result of
the first. Figure~\ref{fig:DisplacedJetsVertex} shows the result, with the four sinusoids in parameter space
corresponding to the four vertices in the original $x-y$ space.

\section{Conclusion}

Enhancing the trigger with massively parallel algorithms offers the possibility of triggering on topological
signatures that were previously prohibitively computationally expensive to reconstruct, thus allowing us to
search for a wide variety of new physics models. The Hough transform, implemented on GPUs using
general-purpose GPU computing, offers a way to improve the tracking reconstruction used in the HLT. This both
helps to counter the effects of increasing pileup at the LHC, by reconstructing all of the tracks in a single
step, and enables us to search for new physics models which are currently not possible or severely limited by
the trigger constraints. While much work remains to test this implementation in more realistic models, the
preliminary results presented here show that this is indeed a promising approach.

\bibliographystyle{iopart-num}
\bibliography{plujan-CHEPProceedings}

\end{document}